\documentclass[12pt,letterpaper]{article}
\usepackage[dvips,usenames]{color}
\usepackage[american]{babel}
\usepackage[latin1]{inputenc}
\usepackage[T1]{fontenc}
\usepackage{xspace}
\usepackage{graphicx}
\usepackage{latexsym}
\usepackage{amsmath,amsfonts,amstext,amssymb,amsbsy,amsopn,amsthm,eucal}
\usepackage{yfonts}[1998/10/03]
\usepackage{dsfont}
\usepackage[dvips,body={8in,11in},vmargin=2cm,hmargin=2cm,head=1cm]{geometry}
\usepackage{url}
\newcommand\email{\begingroup \urlstyle{tt}\Url}
\urldef{\het}\url{www.het.brown.edu}
\urldef{\danieldf}{\email}{danieldf@het.brown.edu}
\usepackage[normalem]{ulem}
\newcommand{\cont}[1]{\ensuremath{\text{\rsfs C\/}^{#1}\xspace}}
\newcommand{\Matrix}[2]{\ensuremath{\text{\rsfs M}_{#1}(\mathbb{#2})}\xspace}
\newcommand{\Cliff}[3]{\ensuremath{\text{\rsfs C}_{#1,#2}(\mathbb{#3})}\xspace}
\newcommand{\ctor}{\ensuremath{\text{\rsfs C}_{3,1}(\mathbb{R})}\xspace}

\newcommand{\mfr}{\ensuremath{\text{\rsfs M}_{4}(\mathbb{R})}\xspace}

\DeclareMathOperator{\diag}{diag}

\DeclareMathOperator{\dop}{d\!}

\newtheorem*{conjec}{Conjecture}

\newcommand{\paslash}{\ensuremath \raisebox{0.025cm}{\slash}\hspace{-0.25cm}\partial\/}
\newcommand{\dfrakslash}{\ensuremath \slash\hspace{-0.20cm} \mathfrak{d}\/}
\newcommand{\Dslash}{\ensuremath \raisebox{0.025cm}{\slash}\hspace{-0.32cm} D}
\newcommand{\dslash}{\not{\hbox{\kern-2pt $\partial$}}}
\newcommand{\pslash}{\not{\hbox{\kern-2.3pt $p$}}}
 \newtoks\nslashfraction
 \nslashfraction={.13}
 \newcommand{\nslash}[1]{\setbox0\hbox{$ #1 $}
   \setbox0\hbox to \the\nslashfraction\wd0{\hss \box0}/\box0 }

\newcommand{\plpl}{\raise-2pt\hbox{$\raise3pt\hbox{$_+$}\hskip-6.67pt\raise0.0pt
  \hbox{$^+$}\hskip 0.01pt$}}
\newcommand{\mimi}{\raise-2pt\hbox{$\raise3pt\hbox{$_-$}\hskip-6.67pt\raise0.0pt
  \hbox{$^-$}\hskip 0.01pt$}}
\newcommand{\bo}{\raise-1mm\hbox{\Large$\Box$}}              
\newcommand{\pa}{\partial}                                       
\newcommand{\dg}{\sp\dagger}                                     
\newcommand{\trans}[1]{{#1}^{\ensuremath{\mathsf{T}}}}           
\newcommand{\hc}[1]{{#1}^{\dg}}                            
\newcommand{\nTH}{{\raise.2ex\hbox{$\displaystyle \bigodot$}\mskip-4.7mu \llap H \;}}
\newcommand{\ket}[1]{\left| #1\right\rangle}              
\newcommand{\ev}[1]{\left\langle #1\right\rangle}        
\newcommand{\inner}[2]{\left\langle #1 , #2\right\rangle}    
\newcommand{\ip}[2]{\left\langle #1 | #2\right\rangle}    
\newcommand{\ipop}[3]{\left\langle#1\left|#2\right|#3\right\rangle} 
\newcommand{\abs}[1]{\left| #1\right|}                    

\newcommand{\comm}[2]{\left[\,#1,#2\,\right]}                 
\newcommand{\acomm}[2]{\left\{\,#1,#2\,\right\}}              
\newfont{\rsfs}{rsfs10.tfm scaled 1200}
\newfont{\bigrsfs}{rsfs10.tfm scaled 2000}
\usepackage{fancyhdr}
\fancypagestyle{plain}{%
\fancyhf{} 
\fancyfoot[C]{\bfseries\thepage} 

}
\newcommand{\helv}{%
\fontfamily{phv}\fontseries{b}\fontsize{9}{11}\selectfont}
\begin{document}
%
%
%
%
\begin{titlepage}
  \begin{flushright}
    {BROWN-HET-1307 \\ March 2002} \\
    \texttt{math-ph/0204024}
  \end{flushright}
  \bigskip
  \begin{center}
  {\Large \textbf{Geometric Calculus and the Fibre Bundle description of Quantum Mechanics}}
  \vspace{1.5cm}\\
  {\large D. D. Ferrante\footnote{email: \danieldf \\ \hspace*{.58cm} www: \het}}
  \vspace{.5cm}\\
  {Department of Physics} \\
  {Brown University, Providence --- RI. 02912}
  \end{center}
  \date{\today}
  \begin{abstract}
    This work has the purpose of applying the concept of Geometric Calculus (Clifford
    Algebras) to the Fibre Bundle description of Quantum Mechanics as done by
    \cite{fbfqm01}. Thus, it is intended to generalize that formulation to curved spacetimes
    [the base space of the fibre bundle in question] in a more natural way.
  \end{abstract}
  \bigskip
  \tableofcontents
\end{titlepage}
\section{Summary of the Fibre Bundle Formulation of Non- and Relativistic Quantum
Mechanics \cite{fbfqm01,fbfrqm02}}
\subsection{Bundle Quantum Mechanics \cite{fbfqm01}}
The mathematical basis for the reformulation of non-relativistic quantum mechanics in
terms of fibre bundles is given by Schrödinger's equation,
\begin{equation*}
  i\, \hslash\, \frac{\dop\psi(t)}{\dop t} = \mathcal{H}(t)\, \psi(t) \; ;
\end{equation*}
$\psi$ is the system's state vector in a suitable Hilbert space $\mathcal{F}$ and
$\mathcal{H}$ is its Hamiltonian.

In the bundle description, one has a Hilbert bundle given by $(F,\pi,\mathcal{M})$, where
the total space is $F$, the projection is $\pi$, the base manifold is $\mathcal{M}$ and a
typical fibre, $\mathcal{F}$,  which is isomorphic to $F_x = \pi^{-1}(x), \; \forall
x\in\mathcal{M}$. Thus, $\exists\; l_x: F_x\rightarrow\mathcal{F}, \; x\in\mathcal{M}$,
isomorphisms. A state vector, $(\psi)$, and the Hamiltonian, $(\mathcal{H})$, are
represented respectively by a state section along paths, $\Psi:
\gamma\rightarrow\Psi_{\gamma}$, and a bundle Hamiltonian (morphisms along paths)
$\mathcal{H}: \gamma\rightarrow\mathcal{H}_{\gamma}$, given by:
\begin{align*}
  \Psi_{\gamma} &: t\mapsto \Psi_{\gamma}(t) = l^{-1}_{\gamma(t)}(\psi(t)) \; ;\\
  \mathcal{H}_{\gamma} &: t\mapsto \mathcal{H}_{\gamma}(t) = l^{-1}_{\gamma(t)}\circ
    \mathcal{H}(t)\circ l_{\gamma(t)} \; ;
\end{align*}
where $\gamma: I\rightarrow\mathcal{M}$, $I\subseteq\mathbb{R}$ is the world-line for some
observer. The bundle evolution operator is given by,
\begin{align*}
  U_{\gamma}(t,s) &= l^{-1}_{\gamma(t)}\circ \mathcal{U}(t,s)\circ l^{-1}_{\gamma(s)} :
    F_{\gamma(s)}\rightarrow F_{\gamma(t)} \; \\
  \Psi_{\gamma}(t) &= U_{\gamma}(t,s)\, \Psi_{\gamma}(s) \; .
\end{align*}

Therefore, in order to write down the bundle Schrödinger equation, $(D\, \Psi = 0)$, a
derivation along paths, $(D)$, corresponding to $U$ is needed,
\begin{equation*}
  D : \text{PLift}^1(F,\pi,\mathcal{M}) \rightarrow \text{PLift}^0(F,\pi,\mathcal{M})\; ;
\end{equation*}
where $\text{PLift}^k(F,\pi,\mathcal{M}) = \{\lambda\; \text{lifting} :
\lambda\in\text{\rsfs C\/}^k\}$, is the set of liftings from $\mathcal{M}$ to $F$, and,
\begin{align}
  \nonumber
  \lambda &: \gamma \rightarrow \lambda_{\gamma}, \;
    \lambda\in\text{PLift}^1(F,\pi,\mathcal{M}) \; ; \\
  \label{eq:pathd}
  D^{\gamma}_s(\lambda) &= \lim_{\epsilon\to 0}
    \frac{U_{\gamma}(s,s+\epsilon)\lambda_{\gamma}(s+\epsilon) -
    \lambda_{\gamma}(s)}{\epsilon} \; \\
\intertext{where}
  \nonumber
  D^{\gamma}_s(\lambda) &= \bigl[(D\lambda)(\gamma)\bigr](s) = (D\lambda)_{\gamma}(s) \; ;
\intertext{and, in local coords,}
  D^{\gamma}_s(\lambda) &= \biggl(\frac{\dop \lambda_{\gamma}^a(s)}{\dop s} +
    \Gamma^a_b(s;\gamma)\, \lambda^b_{\gamma}(s)\biggr)\, e_a(\gamma(s)) \; ;
\end{align}
where $\{e_a(\gamma(x))\}, s\in I$ is a basis in $F_{\gamma(s)}$. Thus, one can clearly
see what is happening with this structure, namely the bundle evolution transport is giving
origin to the linear connection by means of:
\begin{align*}
  \Gamma_a^b(s;\gamma) &= \frac{\pa\bigl(U_{\gamma}(s,t)\bigr)_a^b}{\pa t}\Bigg\vert_{s=t} =
    -\frac{\pa\bigl(U_{\gamma}(t,s)\bigr)_a^b}{\pa t}\Bigg\vert_{t=s} \; ; \\
  U_{\gamma}(t,s)\, e_a(\gamma(s)) &= \sum_b \bigl(U_{\gamma}(s,t)\bigr)^b_a\,
    e_b(\gamma(t))\; ;
\end{align*}
are the local components of $U_{\gamma}$ in $\{e_a\}$.

In this manner, there is a bijective correspondence between $D$ and the bundle Hamiltonian,
\begin{align*}
  \mathbf{\Gamma}_{\gamma}(t) &= \bigl[\Gamma^b_a(s;\gamma)\bigr] =
    \frac{i}{\hslash}\, \mathbf{H}_{\gamma}(t) \; ; \\
  \mathbf{H}_{\gamma}(t) &= i\, \hslash\, \frac{\pa \mathbf{U}_{\gamma}(t,t_0)}{\pa t}\,
    \mathbf{U}^{-1}_{\gamma}(t,t_0) = \frac{\pa \mathbf{U}_{\gamma}(t,t_0)}{\pa t}\,
    \mathbf{U}_{\gamma}(t_0,t) \; ;
\end{align*}
where $\mathbf{H}_{\gamma}$ is the matrix-bundle Hamiltonian.
\subsection{Bundle Relativistic Quantum Mechanics \cite{fbfrqm02}}
\subsubsection{Time-dependent Approach}
The framework developed so far can be generalized to relativistic quantum theories in a
very straightforward way. However, in doing so, it is seen that time plays a privileged
role (thus, leaving the relativistic covariance implicit). Proceeding in such a manner, it
is found that,
\begin{align}
  \label{eq:diracsch}
  i\, \hslash\, \frac{\pa\psi}{\pa t} &= {}^D\!\mathcal{H}\,\psi \; ; \\
  \nonumber
  \psi &= \trans{(\psi_1,\psi_2,\psi_3,\psi_4)} \; ;
\end{align}
where ${}^D\!\mathcal{H}$ is the Dirac Hamiltonian (Hermitian), in the space $\mathcal{F}$
of state spinors $\psi$. Once \eqref{eq:diracsch} is a first-order differential equation,
a \emph{[Dirac] evolution operator}, $({}^D\mathcal{U})$, can be introduced. It is a
$4\times 4$-integral matrix operator uniquely defined by the initial-value problem,
\begin{align*}
  i\, \hslash\, \frac{\pa \psi}{\pa t} {}^D\mathcal{U}(t,t_0) &= {}^D\!\mathcal{H}\circ
    {}^D\mathcal{U}(t,t_0) \\
  {}^D\mathcal{U}(t_0,t_0) &= \mathds{1}_{\mathcal{F}} \; .
\end{align*}

Thus, the formalism developed earlier can be applied to Dirac particles. The spinor
lifting of paths has to be introduced and the \emph{Dirac evolution transport} [along a
path, $\gamma$] is given by,
\begin{equation*}
  {}^D{U}(t,s) = l^{-1}_{\gamma(t)}\circ {}^D\mathcal{U}(t,s)\circ l_{\gamma(s)}, \quad
    s,t\in I \; .
\end{equation*}
The \emph{bundle Dirac equation} is given by (see \eqref{eq:pathd}),
\begin{equation*}
  {}^{D}\!D_{t}^{\gamma}\Psi_{\gamma} = 0 \; .
\end{equation*}
\paragraph{\uwave{Klein-Gordon Equation}}
The spinless, scalar wavefunction $\phi\in\cont{k}, \; k\geqslant 2$, over spacetime,
satisfies the Klein-Gordon equation if (particle of mass $m$, electric charge $e$ and in
an external electromagnetic field given by $A_{\mu}=(\varphi, \vec{A})$),
\begin{equation*}
  \Biggl[\biggl(i\, \hslash\, \frac{\pa}{\pa t} -e\, \varphi\biggr)^2 - c^2\,\biggl(\vec{p} -
    \frac{e}{c}\,\vec{A}\biggr)^2\Biggr]\, \phi = m^2\, c^4\, \phi \; .
\end{equation*}
In order to solve this, a trick can be used. Just let $\psi = \trans{\Bigl(\phi +
  \tfrac{i\,\hslash}{m\,c^2}\, \tfrac{\pa\phi}{\pa t}, \phi - \tfrac{i\,\hslash}{m\,c^2}\,
  \tfrac{\pa\phi}{\pa t}\Bigr)}$. This is a particular good choice if one is interested in
the non-relativistic limit. After some first-order (Schrödinger-type) representation of
the Klein-Gordon equation is chosen, the bundle formalism can be applied to spinless
particles.

Thus, the goal is to describe the given equation of motion in terms of some
Schrödinger-type operator and then apply the bundle formalism \emph{mutatis mutandis}.
\subsubsection{Covariant Approach}
Now, it will be developed an appropriate covariant bundle description of relativistic
quantum mechanics. The difference between the time-dependent and the covariant formalism
is analogous to the one between the Hamiltonian and the Lagrangian approaches to
relativistic wave equations.
\paragraph{\uwave{Dirac Equation}}
The covariant Dirac equation (for a spin $\tfrac{1}{2}$, mass $m$ and charge $e$ particle,
in an external electromagnetic field $A_{\mu}$) is given by,
\begin{align*}
  \bigl(i\,\hslash\, \Dslash &- m\,c\,\mathds{1}_{4\times 4}\bigr)\, \psi = 0 \; ; \\
  \Dslash &= \gamma^{\mu}\, D_{\mu} \; ; \\
  D_{\mu} &= \pa_{\mu} - \frac{e}{i\, \hslash\, c}\, A_{\mu} \; .
\end{align*}
Since it is a first-order differential equation, it admits an \emph{evolution operator}
$\mathcal{U}$, whose job is to connect different values at different spacetime
points. Thus, for $x_1, x_2 \in M_0$, ($M_0$ being the Minkowski spacetime), 
\begin{equation*}
  \psi(x_2) = \mathcal{U}(x_2,x_1)\, \psi(x_1) \; ;
\end{equation*}
where $\mathcal{U}(x_2, x_1)$ is a $4\times 4$-matrix operator, defined as the unique
solution to the initial-value problem:
\begin{align*}
  \bigl(i\,\hslash\, \Dslash - m\, &c\, \mathds{1}_{4\times 4}\bigr)\, \mathcal{U}(x,x_0) =
    0 \; ; \\
  \mathcal{U}(x_0,x_0) &= \mathds{1}_{\mathcal{F}}, \quad x,x_0 \in M_0 \; ;
\end{align*}
where $\mathcal{F}$ is the space of 4-spinors.

Assume that $(F, \pi, \mathcal{M})$ is a vector bundle with total space $F$, projection
$\pi: F\rightarrow \mathcal{M}$, fibre $\mathcal{F}$ and isomorphic fibres $F_x =
\pi^{-1}(x), \; x\in \mathcal{M}$. Then, there exists linear isomorphisms $l_x: F_x
\rightarrow \mathcal{F}$ --- which are assumed to be \textbf{diffeomorphisms} --- so that
$F_x = l_x^{-1}(\mathcal{F})$ are 4-dim vector spaces.

A \cont{1} section\footnote{$\Psi$ is simply a \emph{section} at this time, as opposed to
  the previous two cases, in which it was a \emph{section along paths}. This corresponds
  to the fact that quantum objects do \emph{not} have trajectories in a classical sense.}
is assigned to a state spinor, $\psi$, i.e., $\Psi \in \text{Sec}^1(F,\pi,\mathcal{M})$,
in the following manner:
\begin{equation*}
  \Psi(x) = l^{-1}_x\bigl(\psi(x)\bigr) \, \in \, F_x = \pi^{-1}(x), \; x\in\mathcal{M} \; .
\end{equation*}
Thus, it follows that:
\begin{align*}
  \Psi(x_2) &= U(x_2,x_1)\, \Psi(x_1) \; , \; x_1,x_2 \in \mathcal{M}\; ; \\
  U(y,x) &= l^{-1}_y\circ \mathcal{U}(y,x)\circ l_x \; : \; F_x \rightarrow F_y\; ; \;
    x,y\in\mathcal{M} \; ; \\
  \underbrace{\mathcal{U}(x_3,x_1)}_{\substack{4\times 4 \text{ matrix}\\ \text{operator}} } &=
    \mathcal{U}(x_3,x_2)\circ \mathcal{U}(x_2,x_1) \; ,\; x_1,x_2,x_3 \in \mathcal{M}\; ;
    \\
  \therefore\; U(x_3,x_1) &= U(x_3,x_2) \circ U(x_2,x_1)\; ,\; x_1,x_2,x_3 \in \mathcal{M}
    \; ; \\
  U(x,x) &= \mathds{1}_{F_x} \; , \; x\in\mathcal{M} \; .
\end{align*}
The map $U:\; (y,x)\mapsto U(y,x)$ --- called the \emph{Dirac evolution transport} --- is
a \emph{linear transport along the identity map}, $(\mathds{1}_{\mathcal{M}})$, of
$\mathcal{M}$ in the bundle $(F,\pi,\mathcal{M})$. Thus, it is not difficult to see that
\cite{bp-LTP-general, bp-LTP-appl, bp-LTP-Cur+Tor, bp-LTP-metrics, bp-LTP-Cur+Tor-prop,
  bp-TP-general, bp-TP-parallelT, bp-TP-morphisms, bp-TM-general},
\begin{align*}
  \mathcal{D}_{\mu}\, \Psi &= 0 \; , \;\; \mu=0,1,2,3 \; ; \\
  \mathcal{D}_{\mu} &= \mathcal{D}_{x^{\mu}}^{\mathds{1}_{\mathcal{M}}} \; .
\end{align*}
These are called \emph{Bundle Dirac Equations}.

At this point, local basis could be introduced and a local view of the above could be
written down. However, from this knowledge, what will be of future interest, will come
from the facts,
\begin{align}
  \nonumber
  G^{\mu} &= l^{-1}_x\circ \gamma^{\mu}\circ l_x \; ; \\
  \label{eq:gelcliff}
  \acomm{G^{\mu}}{G^{\nu}} &= 2\, \eta^{\mu\nu}\, \mathds{1}_{\mathcal{F}} \; ;
\end{align}
where $\eta^{\mu\nu}$ is the Minkowski metric tensor, $[\eta^{\mu\nu}] =
\diag(+1,-1,-1,-1)$. This last equation, \eqref{eq:gelcliff}, is the bundle generalization
of \eqref{eq:cliff}. It is clear that this expression can be put in terms of local 
coordinates, in which case it would reduce to (in the equation that follows,
\textbf{boldface} denotes the matrix --- i.e., the expression in local coordinates ---  of
the operator denoted by the same (kernel) symbol),
\begin{equation}
  \label{eq:localcliff}
  \acomm{\mathbf{G}^{\mu}}{\mathbf{G}^{\nu}} = 2\, \eta^{\mu\nu}\, \mathds{1}_{4 \times 4}
    \; ;
\end{equation}
where $\eta^{\mu\nu}$ is the Minkowski metric tensor and $\mathds{1}_{4 \times 4} =
\diag(1,1,1,1)$ is the unit matrix in 4-dim\footnote{The generalization to an arbitrary
  number of dimension is quite clear and straightforward from the equations given.}. Thus,
\eqref{eq:localcliff} is the local expression of \eqref{eq:gelcliff} and a generalization
of \eqref{eq:cliff}.
\bigskip
\section{Spin Manifolds \& Bundles \cite{gp01}}
%
%
It is quite clear how one goes about defining spinors for Minkowski's [flat]
spacetime. The concern in question at this point is how one would do the same for an
arbitrary [curved] manifold \cite{gp01}. In order to do so, let's consider that the
spacetime manifold (which is being assumed space- and time-orientable), $\mathcal{M}$, can
be patched $\mathcal{P}=\{U_1, U_2, \dotsc\}$, and that $\forall U_i \in \mathcal{P},
(i=1, 2, \dotsc), \exists\, \mathcal{V}=\{e_{U_1}, e_{U_2}, \dotsc\}$:
\begin{equation}
  \label{eq:cliff}
  \inner{\mathbf{e}^U_{\mu}}{\mathbf{e}^U_{\nu}} = \eta_{\mu\nu} \; ,  
\end{equation}
where $\mathcal{V}$ is also known as the ``vierbein'' and $\eta_{\mu\nu}=\diag(-1, +1, +1,
+1)$. If it is possible to find Lorentz's matrices, $\Lambda(x) \in Sl(2,\mathbb{C})$, (as
opposed to its Minkowski's counterpart, $\Lambda$, which did not depend on any specific
spacetime point) that will work as [continuous] transition functions for the vierbein
above, then it is said that the structure group of the tangent bundle $\mathcal{M}$ was
\emph{lifted} from the Lorentz group to the group $Sl(2,\mathbb{C})$ and that
$\mathcal{M}$ has a \textbf{spin structure}. The point of having a \emph{spin structure}
is that one can decide --- upon transportation of any frame around a closed path, $C$, in
$\mathcal{M}$ --- whether the frame has made an even or odd number of rotations. The
physical consequence would be the following: Assume that $\mathcal{M}$ has a spin
structure (for, otherwise, the Dirac equation cannot be considered). Say, for example,
that $\mathcal{M}=M_0$ is Minkowski's spacetime. Thus, the Dirac spinors defined over
$M_0$ will be cross sections of a bundle over $M_0$ associated to the tangent bundle
through the representation of $Sl(2,\mathbb{C})$ via $Gl(4,\mathbb{C})$ \footnote{This
  representation is, namely, given by $\rho : Sl(2,\mathbb{C}) \rightarrow
  Gl(4,\mathbb{C})$, $\rho(A) = \diag(A, (\hc{A})^{-1})$.}. Thus, these spinors have the
property that a complete $\bigl[\mathbb{R}^3\bigr]$ rotation will take them to their
negative. The \emph{obstruction} to having a spin structure is measured by the cohomology
groups of $\mathcal{M}$. For example, a spin structure exists if
$H_2(\mathcal{M},\mathbb{Z}_2)$ vanishes, where $H_2(\mathcal{M},\mathbb{Z}_2)$ is the
second homology group with $\mathbb{Z}_2$ coefficients. Thus, if $\mathcal{M}$ does have a
spin structure, the Lorentz structure group can be replaced by $Sl(2,\mathbb{C})$; the
fiber for the tangent bundle of $\mathcal{M}$ is still $\mathbb{R}^4$. If the $c_{UV}$ are
the Lorentzian transition functions for the tangent bundle, $c'_{UV}$ shall be its
$Sl(2,\mathbb{C})$ counterpart. It is, then, constructed the \emph{new}, 4-component,
\textbf{Dirac Spinor Bundle}, $\text{\rsfs S} = \text{\rsfs S}(\mathcal{M})$, whose fiber
is $\mathbb{C}^4$ and transition functions are
\begin{align*}
  \rho_{UV} &: \,U\cap V \longrightarrow Gl(4,\mathbb{C}) \\
  &\hspace{.3cm} c'_{UV}(x) \longmapsto 
    \begin{bmatrix}
      c'_{UV}(x) & 0\\
      0 & \Bigl(\hc{c'}_{UV}(x)\Bigr)^{-1}
    \end{bmatrix} \; .
\end{align*}
This Dirac spinor bundle, ({\rsfs S}\/), is simply the \emph{vector bundle associated to the
$Sl(2,\mathbb{C})$ tangent bundle via the representation $\rho$.} This is the bundle
whose sections, $\psi$, will serve as wave ``functions'' on $\mathcal{M}$. The only
ingredient missing, at this point, is a \textbf{spin connection} in $\text{\rsfs
S}(\mathcal{M})$. The \emph{spin connection} in $\text{\rsfs S}(\mathcal{M})$ is given
by,
\begin{align*}
  \Omega &= \frac{1}{4}\, \omega^{\mu\,\nu}\, \gamma_{\mu}\, \gamma_{\nu}\; ; \\
  &= \frac{1}{8}\,\omega_{\mu\,\nu}\, \comm{\gamma^{\mu}}{\gamma^{\nu}} \; ;
\end{align*}
where $\omega = (\omega_{\mu\,\nu})$ is the Levi-Civita connection for the
pseudo-Riemannian manifold, $\mathcal{M}$, $\gamma_{\mu}$ are Dirac's
$\gamma$-matrices and it was used that $\omega_{\mu\,\nu} =
-\omega_{\nu\,\mu}$. Thus, the \emph{covariant derivative} in the spinor bundle and
the \emph{curved} Dirac operator on $\psi$ are, respectively, given by:
\begin{align*}
  \nabla_t\psi &= \frac{d\psi}{dt} + \frac{1}{4}\, \omega_{\mu\,\nu}\,
    \biggl(\frac{dx}{dt}\biggr)\, \gamma^{\mu}\, \gamma^{\nu}\, \psi \; ;\\
  \Dslash \psi &= \paslash\psi + \frac{1}{4}\, \omega_{\rho\,\nu}^{\mu}\, \gamma^{\rho}\,
    \gamma_{\mu}\, \gamma^{\nu}\, \psi \; .
\end{align*}
\bigskip
\section{Geometric Calculus and the Ordering Problem in Quantum Mechanics
  \cite{gcroaqtcs01, cagc84, isg89}}
%
%
Given the motivation proposed by \cite{gcroaqtcs01}, one can use the Geometric Calculus
in order to prevent the ordering problem that haunts quantum mechanics. Let's make a quick
overview of the method. Given non-commuting numbers, $\gamma_{\mu}\in \ctor\simeq\mfr$,
(where $\ctor$ is a real Clifford Algebra\footnote{The reader should be aware of the
  adopted notation, where, for a metric $g$, its signature is given by $p$ plus and $q$
  minus signs, with $p+q=n$. The structure of the Clifford Algebra can only depend on $p$
  and $q$, thus it is denoted $\Cliff{p}{q}{K}$, where $\mathbb{K}$ is a field (usually
  taken to be either $\mathbb{R}$ or $\mathbb{C}$). Also, it is a well-known
fact that, $\dim\bigl(\Cliff{p}{q}{K}\bigr)=2^n$.} and $\mfr$ is the space of $4\times 4$
real matrices) it is known that

\begin{equation}
  \label{eq:cliffalg}
  \acomm{\gamma_{\mu}}{\gamma_{\nu}} = \gamma_{\mu}\, \gamma_{\nu} + \gamma_{\nu}\,
    \gamma_{\mu} = 2\, g_{\mu\nu} \; .
\end{equation}
(It should be noted that $\Cliff{2}{0}{R} \simeq \Cliff{1}{1}{R} \simeq
\Matrix{2}{R}$. Thus, one can choose to work either with a 4- or a 2-spinor, depending
only on the choice of the Clifford Algebra desired\footnote{This is a subtle point for
 Quantum Mechanics, given that, once pure spinors are not part of that theory, the type [of
 spinor] that one wants to introduce in the theory is completely arbitrary.}.)

Thus, one can introduce the expansion of an arbitrary vector and the dual basis as
follows\footnote{The reader should note that the common ``slash''-notation is not being
  used here. This should not be a source of [future] confusion.}:
\begin{align*}
  \underline{v} &= \underline{v}^{\mu}\,\gamma_{\mu}\; ; \\
  \acomm{\gamma^{\mu}}{\gamma^{\nu}} &= g^{\mu\nu} \; ; \\
  \gamma^{\mu} &= g^{\mu\nu}\, \gamma_{\nu} \; .
\end{align*}

In the same fashion \cite{cagc84, isg89},
\begin{align*}
  \pa &\equiv \gamma^{\mu}\, \pa_{\mu}\; ; \\
  \pa_{\mu}\, \gamma_{\nu} &= \Gamma^{\alpha}_{\mu\nu}\, \gamma_{\alpha}\; ; \\
  \pa_{\mu}\, \gamma^{\nu} &= -\Gamma^{\nu}_{\mu\alpha}\, \gamma^{\alpha}\; ; \\
  \pa\,\underline{v} &= \gamma^{\mu}\,\gamma^{\nu}\,D_{\mu}\,\underline{v}_{\nu} \; ; \\
  \therefore\quad \pa\,\pa\,\psi &= \gamma^{\mu}\,\gamma^{\nu}\,D_{\mu}\,D_{\nu}\, \psi \;
    ;
\end{align*}
where $\Gamma^{\alpha}_{\mu\nu}$ is the \emph{connection}, $\psi\in\mathbb{K}$ is a
scalar and $D_{\mu} = (\pa_{\mu} - \Gamma^{\alpha}_{\mu\nu})$ is the \emph{covariant}
derivative. If the connection is symmetric (vanishing torsion), then
\begin{equation}
  \label{eq:spinmotion}
  \pa\,\pa\,\psi = D_{\mu}\, D^{\mu}\, \psi \; ;
\end{equation}
which is just D'Alembert's operator in curved spacetime.

Thus, one has that \cite{gcroaqtcs01} ($\hslash = 1$),
\begin{equation}
  \label{eq:popentum}
  \mathfrak{p} \equiv -i\, \pa = -i\, \gamma^{\mu}\, \pa_{\mu} \; ,
\end{equation}
which is Hermitian, $(\ev{\mathfrak{p}} = \langle\hc{\mathfrak{p}}\rangle)$, and whose
expectation value, $(\ev{\mathfrak{p}})$,  follows a \emph{geodesic} trajectory in our
curved spacetime.
\bigskip
\section{Putting it all together}
Basically, one wants to generalize eqs \eqref{eq:spinmotion} and \eqref{eq:popentum} to
the bundle formalism previously formulated. In order to do so, let's remember that
($\hslash = 1$, and bringing the ``slash'' notation back):
\begin{align*}
\intertext{\textbf{Bundle} results:}
  G^{\mu} &= l^{-1}_x\circ \gamma^{\mu}\circ l_x \; ; \\
  \mathfrak{d}_{\mu} &= l^{-1}_x\circ \pa_{\mu}\circ l_x \; ; \\
  \dfrakslash &= G^{\mu}(x)\circ \mathfrak{d}_{\mu} \; ; \\
  \dfrakslash &= l^{-1}_x\circ \paslash\circ l_x \; . \\
\intertext{Thus, the bundle version is given by:}
  \mathfrak{p} &= -i\, \dfrakslash = l^{-1}_x\circ (-i\, \paslash)\circ l_x \; ; \\
\intertext{or, in local coords (\textbf{boldface} being matrix-notation, see
\eqref{eq:gelcliff} and \eqref{eq:localcliff}):}
  \mathbf{p} &= -i\, \mathbf{G}^{\mu}\pa_{\mu} \; .
\end{align*}

This works for (non-relativistic) Quantum Mechanics and for Relativistic Quantum
Mechanics.

When dealing with Quantum Field Theories, somethings have to be said before conclusions
are drawn. Let's start with a quick overview of the relevant facts.
\bigskip
\begin{center}
  \uwave{\hspace{15cm}}
\end{center}
    In the functional [Schrödinger's] representation for a \textsc{free scalar} QFT, one
    has that:
    \begin{align*}
      S &= \int \mathcal{L}\, d^4x = \frac{1}{2}\int\bigl(\pa^{\mu}\varphi\pa_{\mu}\varphi -
        m^2\varphi^2\bigr)\, d^4x \; ; \\
    \intertext{where, the conjugate field momentum (note the non-covariant formalism) is:}
      \pi(x) &= \frac{\pa\mathcal{L}}{\pa(\pa_t \varphi)} = \dot{\varphi}(x) \; ; \\
    \intertext{and, the Hamiltonian is:}
      H &= \frac{1}{2}\int\bigl(\pi^2 + \abs{\nabla\varphi}^2 +
      m^2\varphi^2\bigr) \, d^3x \; . \\
    \intertext{The equal-time commutation relations are given by,}
      \comm{\varphi(\vec x,t)}{\pi(\vec y,t)} &= i\delta(\vec{x} - \vec{y}) \; ; \\
      \comm{\varphi(\vec x,t)}{\varphi(\vec y,t)} &= 0 = \comm{\pi(\vec x,t)}{\pi(\vec y,t)}
        \; ; \\
    \intertext{In the coordinate representation, with a basis for the Fock space, where
      $\varphi(\vec x)$ is (now) time independent and diagonal (note that $\phi(\vec x)$ is
      just an ordinary scalar function), one has that:}
      \varphi(\vec x) \ket{\phi} &= \phi(\vec x) \ket{\phi} \; ; \\
      \therefore\; \Psi[\phi] &= \ip{\phi}{\Psi} \; ; \\
      \frac{\delta}{\delta\phi(\vec x)}\phi(\vec y) &= \delta(\vec{x} - \vec{y}) \; \\
      \therefore\; \comm{\frac{\delta}{\delta\phi(\vec x)}}{\phi(\vec{y})} &= \delta(\vec{x} -
        \vec{y})\; . \\
    \intertext{Thus, the functional representation of the equal-time commutators turns out
        to be:}
      \Rightarrow\; \pi(\vec x) &= -i\frac{\delta}{\delta\phi(\vec x)} \; ; \\
      \ipop{\phi'}{\pi(\vec x)}{\phi} &= -i\frac{\delta}{\delta\phi(\vec x)}\delta[\phi' -
        \phi] \; ; \\
    \intertext{and, the momentum operator, $P_i$, which generates spatial displacements, is:}
      \comm{P_j}{\varphi(\vec x,t)} &= -i\frac{\pa}{\pa x^j}\varphi(\vec x,t) \; ; \\
      \therefore\; P_j &= -\int\bigl(\varphi(x)\, \pa_j\pi(x)\bigr)\, d^3x \; ; \\
        P_j &= i\, \int\biggl(\phi(\vec x)\, \pa_j\, \frac{\delta}{\delta\phi(\vec x)}\biggr)\,
        d^3x \; ; \\
    \intertext{thus, using \eqref{eq:popentum}, one has that:}
      \mathfrak{P} &= \gamma^j\, P_j \; . \\
    \intertext{On the other hand, in the momentum representation, $\pi(x)$ is diagonal and time
      independent, which gives:}
      \pi(\vec x) \ket{\varpi} &= \varpi(\vec x)\ket{\varpi} \; ; \\
      \Psi[\varpi] &= \ip{\varpi}{\Psi}\; ; \\
      \varphi(\vec x) &= i\frac{\delta}{\delta\varpi(\vec x)} \; ; \\
        E\, \Psi[\varpi] &= \underbrace{\frac{1}{2}\int\biggl(-\frac{\delta}{\delta\varpi(\vec
        x)}\bigl(-\nabla^2 + m^2\bigr)\frac{\delta}{\delta\varpi(\vec x)} + \varpi^2(\vec
        x)\biggr)\, d^3x}_{= H} \; \Psi[\varpi] \; . \\
    \intertext{Let us introduce a \emph{functional} version of the Fourier transform given by:}
      \Psi[\varpi] &= \int \Psi[\phi]\, e^{i\int\varpi(\vec x)\, \phi(\vec x)\, d^3x} \,
        \mathcal{D}\phi \; . \\
    \intertext{Thus, for a \textsc{free spinor} QFT, analogous relations are valid:}
      H &= \int\hc{\Psi}(x)\bigl( -i\, \gamma^{\mu}\nabla_{\mu} + m\bigr)\Psi(x) \, d^3x \; ; \\
      \acomm{\Psi_{\alpha}(\vec{x},t)}{\hc{\Psi}_{\beta}(\vec{y},t)} &= \delta_{\alpha\beta}\,
        \delta^3(\vec{x} - \vec{y}) \; ; \\
      \acomm{\Psi_{\alpha}(\vec{x},t)}{\Psi_{\beta}(\vec{y},t)} &= 0 =
        \acomm{\hc{\Psi}_{\alpha}(\vec{x},t)}{\hc{\Psi}_{\beta}(\vec{y},t)} \; ; \\
    \intertext{and in the coordinate representation,}
      \Psi(\vec x)\ket{\psi} &= \psi(\vec x)\ket{\psi} \; ; \\
    \intertext{where $\Psi(\vec x)$ is an anticommuting \textsc{field}, thus $\psi(\vec x)$ must
      be a spinor of Grassmann functions $\Rightarrow\; \psi^2_{\alpha}(\vec x)=0$, which
        leads us to:}
      \Phi[\psi] &= \ip{\psi}{\Phi} \; ; \\
      \hc{\Psi}_{\beta}(\vec x) &= \frac{\delta}{\delta\psi_{\beta}(\vec x)} \; ; \\
      \therefore\; E\, \Phi[\psi] &= \underbrace{\int\biggl(\frac{\delta}{\delta\psi(\vec
        x)}\, (-i\,\gamma^{\mu}\nabla_{\mu} + m)\, \psi(\vec x)\biggr)\, d^3x}_{= H}\;
        \Phi[\psi] \; ; \\
      \therefore\;\; \mathfrak{E} &= H\, \gamma^0\; .
\end{align*}

Thus, from all of the above, it is not difficult to see that, a four ``spin-vector'' can
be constructed out of:
\begin{align*}
  \mathfrak{E} &= H\, \gamma^0\; ; \\
  \mathfrak{P} &= P_j\, \gamma^j \; ; \\
  \therefore\;\; \mathcal{P} &= (\mathfrak{E},\mathfrak{P}) = \bigl( H\, \gamma^{0} ,
    P_j\, \gamma^{j} \bigr) \; .
\end{align*}
\begin{center}
  \uwave{\hspace{15cm}}
\end{center}
\bigskip
\begin{align*}
\intertext{In a \textbf{covariant} formulation (the conserved quantity being the
  energy-momentum tensor), one would have that:}
  T^{\mu\nu} &= \frac{\pa \mathcal{L}}{\pa\bigl(\pa_{\nu}
    \varphi_a\bigr)}\pa^{\mu}\varphi_a - g^{\mu\nu}\, \mathcal{L} \; . \\
  \therefore\;\; \bigl( T^{\mu\nu}\, \gamma_{\mu}\, \gamma_{\nu} \bigr) &:
    \text{invariant quantity} \; ; \\
\intertext{thus, using the original notation:}
  \mathcal{T}^{\mu\nu} &= l^{-1}_x\circ T^{\mu\nu}\circ l_x \; ; \\
  \mathfrak{T} &= \mathbf{G}_{\mu}\, \mathcal{T}^{\mu\nu}\, \mathbf{G}_{\nu} =
    l^{-1}_x\circ \bigl( T^{\mu\nu}\, \gamma_{\mu}\, \gamma_{\nu} \bigr)\circ l_x \; ; 
\end{align*}
The reader should note that, the properties described in \cite{gcroaqtcs01} will easily
generalize to the above cases. This leads us to the following thought.
\begin{conjec}
  The basic description of physical quantities should be done in terms of spin-variables,
  such as spinors, spin-vectors and spin-tensors.
\end{conjec}
\bigskip
\section{Aknowledgements}
This work was partly supported by \textsf{DOE} grant \textsf{DE-FG02-91ER40688 - Task D}\/.
\bibliographystyle{hplain}
\bibliography{bqmG}
\end{document}